\documentclass{amsart}

\usepackage{amsmath,amssymb,amsthm,amsfonts,amsrefs}
\usepackage{mathrsfs,mathtools} 
\usepackage{setspace}
\DeclareMathOperator*{\argmin}{arg\,min}
\DeclareMathOperator*{\argmax}{arg\,max}
\usepackage[]{algorithm2e}
\usepackage{verbatim}

\usepackage [pagewise] {lineno}

\usepackage{hyperref}

\newtheorem{thm}{Theorem}[section]

\begin{document}


\title{Estimating on-street parking occupancy using smart meter data}



\maketitle

\author{%
  \noindent \textbf{Daniel Jordon} 
  
  \noindent Independent
  
  \noindent \url{daniel.jordon@gmail.com}
  
    \hfill\break
  \textbf{Robert C. Hampshire}
  
    \noindent Ford School of Public Policy 
  
  \noindent University of Michigan
  
  \noindent Ann Arbor, MI 48109
  
  \noindent \url{hamp@umich.edu}
  
  \hfill\break
  \textbf{Tayo Fabusuyi}, \textit{Corresponding Author}
  
    \noindent Transportation Research Institute 
  
  \noindent University of Michigan
  
  \noindent Ann Arbor, MI 48109
  
  \noindent \url{Fabusuyi@umich.edu}
}


\newpage

\noindent ABSTRACT. The excessive search for parking, known as \textit{cruising}, causes pollution and congestion. To mitigate these negative effects, cities are exploring new methods. However, accurately measuring the number of vehicles searching for parking is challenging and typically requires sensing technologies. In this paper, we propose an alternative method that eliminates the need for such technology by leveraging parking meter payment transactions to estimate parking occupancy and the number of cars searching for parking. Our estimation scheme is based on Particle Markov Chain Monte Carlo (PMCMC).  

We validate the PMCMC approach using data simulated from a $GI/GI/s$ queue, demonstrating that it produces asymptotically unbiased Bayesian estimates of parking occupancy and key model parameters such as arrival rates, average parking times, and payment compliance rates. Finally, we apply our method to parking meter data from SF\textit{park}, a large scale parking experiment and subsequently, compare the Particle Markov Chain Monte Carlo parking occupancy estimates to the actual data from parking sensors. Our approach is easy to replicate and scalable, relying solely on historical parking payment transaction data that cities already have.







\hfill\break

\newpage

\section{Introduction}

Finding an available parking space is an annoyance of modern life. The search for parking has external costs as well given that excess driving increases traffic congestion and pollutes the environment. In response to these negative externalities, city officials are deploying parking information and pricing systems. Parking occupancy information and market based parking pricing strategies reduce the time to find a parking space and pollution (\cites{fabusuyi2014, pierce2013getting, millard2014curb, chatman2014theory}). These strategies require cities to install both parking occupancy sensors and automated payment stations. However, on-street parking sensors are expensive to install and maintain, and many cities cannot afford to deploy parking sensors. In contrast to parking sensors, the automated payment stations are more affordable and more widely deployed. In this paper, we attempt to answer the following question: is it possible to estimate parking occupancy and the number of drivers searching for parking using only parking meter payment data?

\begin{figure}[h] 
\begin{center}
\includegraphics[scale=1]{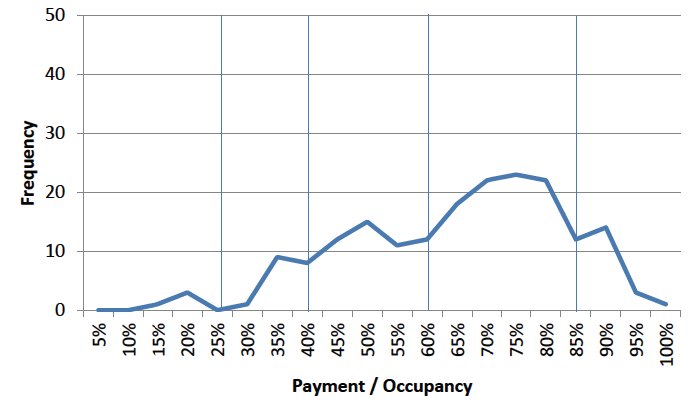}
\end{center} 
\caption{ Probability distribution of ratio of paid parking time to occupied time. We observe many blocks experience a low payment compliance rate. Seventy-five percent of the  occupied time on the modal block is paid for. Source: San Francisco Metropolitan Transportation Authority   } \label{fig:paymentOcc}
\end{figure}

 Two challenges stand in the way of answering this question. First, a large fraction of drivers do not pay at all or do not pay enough to cover their parking time. For example, in San Francisco the non-payment rate is roughly 30 percent, driven in part by the perennial problem with the abuse of disabled parking placards for on-street parking  \cite{SIRA2014}. Figure \ref{fig:paymentOcc} plots the probability distribution of the non-payment rates in San Francisco. During metered hours, only 70\% of parking time is paid for on the modal block. The second challenge is that drivers almost never pay for the exact amount of parking that they need, as they either pay too much or too little.

It is these challenges that are estimation approach seeks to address. Specific contributions of our approach include the:
\begin{itemize}
\item  The introduction of a modeling framework to infer parking occupancy and the time to find parking given a sequence of parking meter payments using a variant of the Particle Markov Chain Monte Carlo (PMCMC) method. We show that this algorithm generates asymptotically unbiased estimates of parking occupancy of the underlying model parameters, such as arrival rates, average parking time, and non-compliance rates (\S \ref{sec:model:pmcmc}).
\item The use of a stochastic queueing model to provide constraints for inferring the unobserved parking occupancy and the model parameters.  (\S \ref{sec:model:parking}).
\item The validation of the performance of the proposed methods with data from a large scale parking experiment called SF\textit{park} (\S \ref{sec:results:sfpark}) and with
simulated data from $GI/GI/s$ queue (\S \ref{sec:results:simulation}).
\end{itemize}

The sample path construction for the $GI/GI/s$ queue is the foundation of the our queueing inference methods. The sample path formulation leads to a flexible modelling framework for inference. Our proposed method extends to time varying model parameters (arrival rates, service rates, etc). For the ease of exposition, we focus on the case of constant model parameters. Furthermore, our methods extends to the case of random order of service. There exists an analogous sample path construction of a GI/GI/s queue with random order of service (see Sutton and Jordon \cite{sutton2011bayesian}). For the ease of exposition, we focus on the first come first served service discipline.

The balance of the paper is organized as follows: Section 2 provides a review of the existing literature. The details of the application of a stochastic queuing model to parking occupancy is discussed in Section 3. Information on the computational algorithm and how the queuing behavior is inferred from smart parking meter transaction data is provided in Section 4. Numerical experiments are carried out in Section 5 and Section 6, the last section, concludes.

\section{Literature Review}

Traffic congestion has a huge negative impact not just on local municipalities but also on the global economy. It increases pollution, wastes energy, and costs billions in lost time and productivity. A significant contributor to traffic congestion is vehicles searching for parking and it is estimated that in busy areas, an average of 30 percent of traffic congestion is from drivers cruising for open parking spots \cite{shoup2006cruising}. Also contributing to this problem is the increase in vehicle ownership \cite{Schaller2021}, a development which adds to the scarcity of parking spaces, particularly, on-street parking. Monitoring on-street parking is a much more difficult task than monitoring garage parking, which is relatively easy to compute through gate counts of entering and exiting vehicles. And cities often price on-street parking significantly lower than off-street parking, leading drivers rationally to choose to cruise \cite{millard2014curb}.

Intelligent transportation systems (ITS) such as smart parking solutions are being used to help relieve such congestion and help drivers navigate this situation by providing real-time and predictive information about current and future traffic and parking availability \cite{fabusuyi2014}. One well-known smart parking solution to the “cruising for parking” problem was SFpark \cite{SIRA2014}. SFpark was a Federal Highway Administration (FHWA) funded parking management project in San Francisco, California from 2009 to 2014. The City of San Francisco installed close to 12,000 magnetometer sensors in 8000 parking spaces, creating a wireless sensor network to collect and distribute information on parking availability [6]. Though conceived as a demonstration project, the problems with replicating the program elsewhere include the enormous cost, which is out of the reach of most cities, the limited life-cycle of the fixed sensors, and the variable pricing mechanism used by SFpark \cite{fabusuyi2018}.

Apart from the demonstration program, other approaches of managing on-street parking include sensing using smartphones \cite{nawaz2013} or through sensors attached to vehicles \cite{mathur2010} which requires drivers to opt into the program. Crowdsensing, via on-vehicle sensors, is another opt-in system since it depends entirely on the willingness of drivers to attach the sensors to their vehicles and to provide information for others to use \cite{Liao2016}. Other forms of sensing include a combination of on-vehicle sensors and crowdsourcing \cite{roman2018}. User opt-in is required in this type of project as well.

Detecting parking occupancy is also carried out via vehicle to infrastructure (V2I) communication. An example is a novel on-street parking management system that detects parking occupancy using low-cost Bluetooth beacon transmitters installed in vehicles that are connected to receivers located near on-street parking spots \cite{Chen2019}. Some of these solutions are being advocated by the private sector or by the private sector in partnership with academia. These include the use of radar sensors mounted on streetlights by Siemens called Integrated Smart Parking Solution \cite{Siemens2015}. Here, the radar sensors are mounted on streetlights to scan a bigger area than can be done from a single vehicle. The radar sensors monitor both traffic flows and parking spots, to help provide information to drivers searching for parking

Ford Motor Company and Georgia Institute of Technology (Georgia Tech) have collaborated to build a system that gives information about on-street parking availability using sonars and radars on existing cars \cite{Ford2015}. This technology is available to Ford drivers as a paid service. Other systems use vehicles’ pre-installed parking sensors to better classify on-street areas into either legal or illegal parking spots in order to give more accurate information to drivers searching for parking \cite{coric2013}. Using data from parking meters or kiosks is yet another method of collecting valuable parking information for drivers given that parking meters or kiosks are responsible for 95 percent of on-street paid parking spots \cite{Yang2017}. These transaction data have been used with other data to provide not just the real time information on parking availability but also to forecast short term on-street parking occupancy. Yang et. al. \cite{Yang2019}, for example, use a deep neural network -based model that uses parking meter transactions along with traffic and weather conditions to predict parking occupancy. Yang et. al. earlier approach \cite{Yang2017} simulated individual payment and parking behavior using a probabilistic payment model, which when accumulated over all transactions, creates time-varying occupancy estimations with a low error rate.

The use of parking kiosks or meters to estimate parking occupancy is however, complicated by the fact that drivers can neglect to pay for parking, pay more than required for their time spent, or pay less than required. We can use the parking meter payments in a queueing process model of on-street parking to create a queueing inference engine (QIE) \cites{larson1990queue, bertsimas1992deducing, mandelbaum1998estimating}. However, our approach builds on the literature in three significant ways. First, our underlying queueing model has no restriction on the inter-arrival time distribution. Second, our parking meter data makes up a subset of service commencement times, in contrast to the QIE whose observations are the complete set of departure times. Additionally, we only observe a noisy estimate of the service times; namely the payment amounts. Finally, we introduce a completely different framework based on simulating the queuing process and Markov chain Monte Carlo methods.

Our approach is innovative given that the authors know of no existing work that applies a QIE approach for estimating on-street parking spaces using parking meter transaction data. Our approach improves on the San Francisco Municipal Transportation Authority \cites{SIRA2014, demisch2016demand} regression-based model for estimating hourly parking occupancy using payment data. First, our model provides predictions at the time scale of observed payments while SFMTA’s regression-based approach is at the hourly level. Secondly, our framework learns the model parameters on the fly, while previous methods require the user to tune the regression approach for each geographic region. Thus, the ease of replicating our approach is afforded users as they can deploy our method more broadly compared to the regression-based approach.

\section{Models}
\subsection{Parking Model} \label{sec:model:parking}

The unit of analysis is a single on-street parking block with $s$ parking spaces. We model parking occupancy at the block level as a $GI/GI/s$ stochastic queueing model. We define $\{\alpha_j\}$ to be a sequence of random independent inter-arrival times of the drivers  with a general distribution and $\{\nu_j\}$ as a sequence of random independent parking (service) times with a general distribution. These inter-arrival and service times are the random input primitives to the queue.  The queue transforms the random input primitives into arrival times, service start times, and departure times. The physics of the queue determine the values of these outputs.  The queue transforms the random primitives according to the set of equations (see \cites{kiefer1955theory,krivulin1994recursive,sutton2011bayesian})

\begin{align}  \label{eq:ggs_queue}
a_j &= a_{j-1} + \alpha_j \nonumber  \\
b_{ji} &= \max\{ d_{j'} | a_{j'} < a_j,p_{j'}=i\} \nonumber \\
c_j &= \min_{i \in (0,s]} b_{ji} \nonumber \\
p_j &= \argmin b_{ji} \nonumber \\
u_j &= \max\{a_j,c_j\}  \nonumber \\
d_j &= \nu_j + u_j 
\end{align}
where the outputs are the sequences of arrival times $\{a_j\}$ and departure times $\{d_j\}$. Intermediate variables assist the transformation;  $u_j$ is the service start time of $j$-th arrival, the $j$-th arrival parks at space $h_j$, and  $c_j$ is the first time a parking space becomes available to serve the $j$-th arrival. This transformation is one-to-one and invertible \cite{sutton2011bayesian}. Additionally, the likelihood of any set of $N$ arrival and departure times is

\begin{equation} \label{eqn:joint_dens}
 p(\mathbf{a},\mathbf{d}) = \prod_{j=1}^N h_\alpha(a_j-a_{j-1}) \cdot h_\nu(d_j-u_j)
\end{equation} 
where $h_\alpha$ is the density of the inter-arrival times with parameters $\theta_1$ and $h_\nu$ is the density of the service times with parameters $\theta_2$. This product form of the likelihood is due to the independence of the inter-arrival times and the service times.

These equations correspond to drivers parking at available spaces in the order they arrive.   We can extend our results to random order of service by slightly modifying the queue input and output equations \cite{sutton2011bayesian}. For ease of presentation, we focus on the first-come-first-served (FCFS) service discipline


Using this transformation, we construct the number of parked cars, the number of cars searching for parking, and the time needed to find an available parking space. 

\subsection{Payment Observation Model} \label{sec:model:payment}
We now introduce payments into the parking model. We assume that a driver pays immediately after parking. The driver pays at an automated payment station that aggregates payments for all $s$ spaces on the block. After the $k$-th driver parks, the time remaining on the meter is
\begin{equation*}
     m_k = \max(m_{k-1} + \beta_k - (\tau_k - \tau_{k-1}) , 0)
\end{equation*}
where $\beta_{k}$ is the amount paid time by the $k$-th parking driver and $\tau_{k}$ is the time of the $k$-th payment. The payment amount relates to the true parking time $\nu$ through a general joint distribution. For the sake of simplicity, we assume that the payment amount is the mixture of two distributions. The first component of the mixture is an exponential distribution with a mean equal to the true parking time. The second component of the mixture is a point mass at zero which represents the probability, $p$, of a driver not paying for parking. 

The observations consist of the time of each payment $\tau_{1:K}$ and the time remaining on the meter $m_{1:K}$.  We focus on analyzing a batch of observations $y_{1:K}= (\tau,m)_{1:K}$.

\subsection{State Space}
An important feature of the model is that not all parking drivers pay for parking. Therefore, the number of payments do not correspond to the number of arriving vehicles. We now introduce a random variable to track the number of arriving vehicles. Let the random variable $J(k)$ be the number of arriving drivers before the $k$-th observed payment. Since the $k$-th payment occurs at time $\tau_k$, the number of arrivals up to the $k$-th payment satisfies
\begin{equation} \label{eqn:arrivals}
\\J(k) =  \argmax_{n} \bigg\{ \sum_{i=J(k-1)+1}^n 1_{\{\beta_i > 0\}}  \le 1 \bigg\}, \quad
J(k) = J(k-1) + x_k
\end{equation} where $x_k$ is a geometric random variable with rate $P(\beta_k=0)$

The number of arrivals before the $k$-th payment, $J(k)$, evolves according to a Markov chain at payment times. It is a counting process with independent increments. The number of arrivals until $k$ payments has a negative binomial distribution with parameter $1-p$ and $k$.

The state space is the total number of arrivals and the corresponding collection of arrival times, service start times, and departures times of all customers in the queue at the time of the $k$-th payment
\begin{equation*}
x_k = (j(k),a_{1:j(k)}, u_{1:j(k)}, d_{1:j(k)}).
\end{equation*}
The state has two components, the number of arrivals before the $k$-th payment and the queueing quantities for these arrivals. The dimension of $x_k$ grows randomly with $k$ due to the unknown number of arrivals between payments.
\section{Inference}
Let $\theta$ be the parameters of the inter-arrival time, service time and payment distributions in the parking and payment models. Our goal is to estimate the 
\begin{equation} \label{eq:post}
p(\theta,x_{1:K} \,| y_{1:K}) = \frac{p_\theta(x_{1:K},y_{1:K}) \cdot p(\theta)}{p(y_{1:K})}
\end{equation}
where $p(\theta)$ is the prior distribution on the model parameters 

\begin{align*}
p_\theta(x_{1:K}, y_{1:K}) &
  = p_\theta(x_{1:K}) \cdot p_\theta(y_{1:K} | x_{1:K}) \\
& = \mu_\theta{(x_1)} \prod_{k=2}^K f_\theta(x_k | x_{k-1}) \cdot \prod_{k=1}^K g_{\theta}(y_k|x_k)
\end{align*}
where the process $\{x_n;n \ge 1\}$ has an initial density $\mu_\theta(\cdot)$ and the payments are independent of each other given the state and have a common density $g_\theta(y|x)$ where $x$ is the state the system. The transition probability is

$$f_\theta(x_k | x_{k-1})= p^{j(k)-j(k-1)-1}(1-p) \cdot \prod_{j=j(k-1)}^{j(k)} h_\alpha(a_j-a_{j-1}) \cdot h_\nu(d_j-u_j).
$$

This joint density $p_\theta({x_{1:K}, y_{1:K}})$ has the form of a Feynman-Kac model. The rich literature on Feynman-Kac models (see Del Moral \cite{del2000branching})  provides the theoretical foundation for the proposed computational algorithms.  

\subsection{Particle Filter}
The $GI/GI/s$ queue transformation constrains the likely values of the random primitives given the payment observations. First, we consider the case where the parameters of the inter-arrival, service time, and payment distributions are known. Our goal is to compute the posterior distribution of the state given payment observations $p_\theta (x_{1:K} | y_{1:K})$.  The traditional Kalman filter \cite{kalman1960new} does not apply to this goal because the dynamics of the state are neither linear nor does it have a Gaussian distribution. The queueing inference engine techniques  \cites{larson1990queue, bertsimas1992deducing, mandelbaum1998estimating} do not apply to the goal either due to the general nature of the distribution of the random primitives.

The solution is to simulate sample trajectories from $p_\theta (x_{1:K} | y_{1:K})$ using Sequential Monte Carlo (SMC), also known as a particle filter \cites{gordon1993novel, doucet2009tutorial}. The particle filter approximates the posterior distribution over state trajectories using $N$ particles. The particle filter proceeds in three steps. First, for each particle we simulate a one-step state transition. Next, we compute an importance weight for each particle based on the likelihood of observing the current payment given the state of the particle. Finally, we  propagate, kill, or replicate each particle in proportion to its importance weight. Particles with larger weights have a higher likelihood of replicating, and those with smaller weights tend to die out. In this way, the surviving particles approximate samples from the posterior distribution $p_\theta (x_{1:K} | y_{1:K})$. For any number of particles, the particle filter produces an unbiased estimate of the posterior distribution \cite{doucet2009tutorial}. It also produces an unbiased estimate of the likelihood of the observations

\begin{equation}
\hat{p}(y_{1:K}) = \prod_{k=1}^K \bigg( \frac{1}{N} \sum_{n=1}^N w_k^{(n)} \bigg)
\end{equation}
where $w_k^{(n)}$ is the importance weight of the $n$-th particle at step $k$. 

The two ingredients of the particle filter are the state transition probabilities and the observation likelihood model. The particle filter requires only the ability to simulate the one step transitions of the state. The observation likelihood model is more complex to compute than the state transitions.

\begin{algorithm}[t] 
\KwData{Payment Observations over time $y_{1:K}$ }
\KwIn{$\theta$, Number of particles $N$, and Effective Sample Size threshold, $N_{ESS}$}
Sample initial state $\tilde{\mathbf{x}_0} \sim p(\mathbf{x}_0|\theta)$\; 
$W_0^{(n)} \gets \frac{1}{N}$ and $L^0(\theta)=1$\;

 \For{$k = 1, \ldots, K$}{ 
     \For{$n = 1, \ldots, N$}{
        Sample $\tilde{\mathbf{x}}^{(n)}_k \sim f(\mathbf{x}_k | \mathbf{x}^{(n)}_{k-1})$ using  Equation \eqref{eq:ggs_queue}\;
        $\tilde{\mathbf{x}}^{(n)}_{1:k} \gets (\mathbf{x}^{(n)}_{1:k-1}, \tilde{\mathbf{x}}^{(n)}_k  )$\;
        Sample $\tilde{y}^{(n,h)}_k \sim g_k(\cdot; \tilde{\mathbf{x}}^{(n)}_k)$ for each $h = 1\ldots,H$ using the sample path equations\;
        $w_k^{(n)} \gets  w_{k-1}^{(n)} \cdot g_\epsilon(y_k)$ using Equation \eqref{eq:obsModel}\;
        $L^{k+1}(\theta) \gets L^{k}(\theta)  \cdot \frac{1}{N}\sum_{n=1}^N w_k^{(n)}$\;
    } 

    $W_k^{(n)} \gets \frac{w_k^{(n)} }{\sum_{n=1}^N w_k^{(n)} }$ for each $n=1,\ldots, N$\; 
    
    \eIf{$\sum_{n=1}^{N} \frac{1}{[w_k^{(n)}]^{2}} \le N_{ESS}$}{ 

        Resample particles with replacement according to a multinomial distribution with weights $\{ W_k  \}$. Let $n^*$ be the resampled index for particle $n$ where $P\{ n^* = l \}= W_k^{(n)}$ for $l=1,\ldots\, N$. \;
        
        For $n=1,\ldots,N$, set   $\mathbf{x}^{(n)}_{1:k}= \tilde{\mathbf{x}}^{n^*}_{1:k}$ and the weights are uniform $w_k^{(n)} =1/N$.  \;
    }{ 

        $\mathbf{x}^{(n)}_{1:k} \gets \tilde{\mathbf{x}}^{(n)}_{1:k}$ for each $n=1,\ldots,N$\;
    }
} 
\KwOut{Distribution of state trajectories over time, $p_\theta(\mathbf{x}_{1:k} | y_{1:k})$, and likelihood estimate $\hat{p}_{\theta}(y_{1:K}) = \prod_{k=1}^K \frac{1}{N} \sum_{n=1}^N w_k^{(n)}$}. 
\caption{Particle Filter with ABC target.}  \label{alg:abc}
\end{algorithm}

\subsection{Approximate Bayesian Computation (ABC) }

In this section, we construct the observation likelihood, $g(y_{k}| \mathbf{x}_k)$,  of the $k$-th observation given the state. The observations consist of the payment time and time remaining on the meter just after the $k$-th payment. The time of the $k$-th payment must correspond to either an arrival time or a departure time of a previous arrival. However, the structure of the departure times is complex and does not preserve arrival order. Further, the departure times depend on residual service times that are difficult to compute for general service distributions. The amount of the $k$-th payment depends on the service time of the corresponding driver who need not be the $k$-th arrival. These factors make the likelihood challenging to compute in closed form, therefore, we simulate an estimate of the likelihood density. 


For each particle we simulate $H$ realizations of the $k$-th payment given the state $x_k$. This simulation is fast due to the sample path construction of the queue and payment models.  A particle receives a high weight if the simulated $k$-th payments are ``close" to the observed $k$-th payment, and receives a low weight otherwise. This is called approximate Bayesian computation (ABC) \cites{marjoram2003markov, beaumont2002approximate}. Jasra  et al. \cite{jasra2012filtering} and  Calvet et al. \cite{calvet2014accurate} successfully use ABC in the particle filtering context. The full algorithm appears in Algorithm \ref{alg:abc}. We approximate the likelihood density function by using a kernel-based distance metric in the observation space

\begin{equation} \label{eq:obsModel}
 \tilde{g}_\epsilon(y_k ) = \frac{1}{{H \cdot \epsilon }}\sum\limits_{h = 1}^H {K\left( {\frac{{{y_k} - \tilde{y}^{(h)}_k}}{\varepsilon }} \right)} 
\end{equation} where $\epsilon$ is the bandwidth of the kernel, a $H$ is the number of simulations.

\subsection{ Particle Marginal Metropolis-Hastings (PMMH)}\label{sec:model:pmcmc}

We turn to jointly estimating both model parameters. We are interested in the parameters of the inter-arrival time distribution ($\lambda$), the service time distribution ($\mu$) and  payment probability ($p$).  We implement the Particle Markov Chain Monte Carlo method (PMCMC) \cite{andrieu2010particle}. This combines two powerful methods: Markov Chain Monte Carlo and the particle filter. Specifically, we use the Particle Marginal-Metropolis Hastings (PMMH) method. The outputs of PMMH are an estimate of the data likelihood, $\hat{p}_\theta(y_{1:K})$, and samples from the  posterior distribution $p(\theta,x_{1:K} \,| y_{1:K}) $. 

The key idea of the approach is that the particle filter generates state proposals within the Metropolis-Hastings algorithm. The particle filter also generates an estimate of the likelihood $p_\theta(y_{1:k})$ which is part of the proposal acceptance probability. We accept joint state and parameters proposals that explore parts of the posterior distribution that are more likely. In essence, we are searching over time evolution of the state space and parameters that are the most likely given the observed sequence of payments.

Here is how it works. The input to the method is a batch of payments observations, $y_{1:K}$.  First, we generate proposed values for the parameters using a Gaussian random walk proposal mechanism. Next, we run the particle filter to generate a proposed state trajectory $x_{1:K}$ under the posterior distribution and we generate a likelihood estimate $\hat{p}_\theta(y_{1:K})$ under the proposed parameters.  We accept the proposed parameters and state trajectory according to an acceptance probability that depends on the likelihood and the parameter proposal distribution. We iterate this process which results in unbiased estimates of joint distribution over the parameters and the state trajectory. The estimates are consistent for any number of particles.    The output of PMMH is a distribution of accepted state proposal-parameter pairs. Each accepted proposal of the MCMC step consists of both a state trajectory and corresponding model parameters. 

We modify the PMMH algorithm by plugging in the ABC particle filter to generate the state proposals. The full algorithm appears in Algorithm \ref{alg:pmmh}. In their discussion of Andrieu et al. \cite{andrieu2010particle}, Cornebise and Peters \cite{cornebise2009comments} advocate this approach when the likelihood is intractable. Our main result shows that the ABC-PMMH methods generates unbiased estimates of the the target joint distribution as the number of particles and simulated ABC observations go to infinity.


\begin{thm}
The ABC-PMMH converges to the correct posterior distribution when the number of particles grows appropriately with the size of the kernel density estimator, that is
\[
\mathop {\lim }\limits_{N \to \infty } \left| {\log p_\theta ^{ABC} - \log p_\theta ^{}} \right| \to 0,
\]
where the number of particles scales with the number of simulated $H$ payments
\[
N = O\left( {{{\left( {\frac{H}{{\log H}}} \right)}^{ - 2/5}}} \right)
\]
and the bandwidth of the kernel density estimator decreases with $H$ according to 
$\varepsilon  = c \cdot {\left( {\frac{{\log H}}{H}} \right)^{1/5}}$.
\end{thm}
\begin{proof}
The Particle Marginal Metropolis-Hasting acceptance probability depends on the likelihood estimate generated by the particle filter 
\[
{p_\theta }\left( {{y_{1:T}}} \right) = \prod\limits_{t=1}^T \bigg( \frac{1}{N} \sum\limits_{n=1}^N { g_\theta \left( {y_t} | {x_t^n} \right) \bigg)} 
\]
where $T$ is the number of steps (payments) and $N$ is the number of particles. This is an unbiased estimator of the likelihood \cite{del2000branching} for any number of particles.  

However, the expression for the density function $g$ is intractable. Therefore, we estimate the density $g$ using a Kernel density estimator with bandwidth $\epsilon$,
\[
{\hat g_\varepsilon }\left( y \right) = \frac{1}{{H \cdot \varepsilon }}\sum\limits_{h = 1}^H {K\left( {\frac{{{y_h} - y }}{\varepsilon }} \right)}
\]
where $(y_h)_{h=1}^{H}$ are simulated from true distribution $g$. The corresponding observation likelihood estimate generated by the particle filter is 
\[\log p_\theta ^{ABC} = \sum\limits_{t = 1}^T {\sum\limits_{n = 1}^N {\frac{1}{{\varepsilon N H}}\sum\limits_{h = 1}^H {K\left( {\tilde y_t^{n,h};{y_t},{\varepsilon ^2}} \right)} } }.
\]
If $ \mathop {\lim }\limits_{N \to \infty } \left| {\log p_\theta ^{ABC} - \log p_\theta ^{}} \right| \to 0 $, then the posterior distribution converges to the correct posterior distribution by applying Theorem 4 of  Andrieu et al. \cite{andrieu2010particle}. 

The difference between the true and approximate log likelihood is bounded above
\begin{align*}
\left| \log p_\theta ^{ABC} - \log p_\theta^{} \right| &
\leq \sum_{t = 1}^T \frac{1}{N} {\sum_{n = 1}^N {\left| {\frac{1}{{H \cdot \varepsilon }}\sum_{h = 1}^H {K\left( {\tilde y_k^{n,h};{y_k},{\varepsilon ^2}} \right) - {g_\theta }\left( {{y_k}\left| {x_k^n} \right.} \right)} } \right|} } \\
& \leq \sum_{t = 1}^T \frac{1}{N} {\sum_{n = 1}^N {\mathop {\sup }\limits_{{y_k}} \left| {\frac{1}{{H \cdot \varepsilon }}\sum_{h = 1}^H {K\left( {\tilde y_k^{n,h};{y_k},{\varepsilon ^2}} \right) - {g_\theta }\left( {{y_k}\left| {x_k^n} \right.} \right)} } \right|} }\\
& \leq T \cdot \delta  \cdot {\left( {\frac{{\log H}}{H}} \right)^{2/5}}.
\end{align*}
The last inequality follows from the uniform convergence results for kernel density estimators.
\end{proof}

\begin{thm} [Fan and Yao \cite{fan2003nonlinear} ]  If the density $g$ has a bound second derivative and the bandwidth is chosen so that $\varepsilon  = c \cdot {\left( {\frac{{\log H}}{H}} \right)^{1/5}}$, then 
\[\mathop {\sup }\limits_y \left| {{{\hat g}_\varepsilon }\left( y \right) - g\left( y \right)} \right| \le \delta  \cdot {\left( {\frac{{\log H}}{H}} \right)^{2/5}}. \]
\end{thm}

We adopt the PMMH method for two reasons. First, the approach allows for flexibility in model specification and only requires the ability to simulate sample paths of the model.  Secondly, PMMH handles state spaces with random dimensions. Our state space is of an unknown dimension due to the unknown number of arrivals between payments. In the literature this problem is called trans-dimensional  \cite{green2003trans}. One way to perform estimation on trans-dimensional problems is the reversible jump Markov chain Monte Carlo (RJMCMC) method  \cite{green1995reversible}. This popular method requires the development of problem specific and often complex proposal moves \cite{karagiannis2013annealed}. There are several papers that use the PMCMC methods in the trans-dimensional problems. For example, PMCMC was used to construct phylogenetic trees \cite{persing2015simulation}, and to estimate of volatility in asset pricing models \cite{bauwens2014marginal}.

\begin{algorithm}[t]
\KwData{Payment Observations over time $y_{1:K}$ }
\KwIn{ Number of particles $N$, Effective Sample Size threshold, $N_{ESS}$, number of Metropolis-Hastings steps $M$ }

\For{$m = 1, \ldots, M$}{

\eIf{ $m = 1 \,\,$}{
    \begin{itemize}
        \item Pick an initial value of parameters: $\mathbf{\theta}(0)$. \;
        \item Run  Algorithm \ref{alg:abc}  targeting $p_{\theta(0)}(\mathbf{x}_{1:K} | y_{1:K})$ and the data likelihood $p_{\theta(0)}(y_{1:K})$. \;
        \item Sample $X_{1:k}(0) \sim \hat{p}_{\theta(0)}(\mathbf{x}_{1:k} | y_{1:K})$ and the likelihood estimate $\hat{p}_{(\theta(0)}(y_{1:K})$. \;
    \end{itemize}
}{
    \begin{itemize}
        \item Sample from parameter proposals: $\theta^* \sim q(\theta | \theta(m-1) )$ \;
        \item Run Algorithm \ref{alg:abc}  targeting $p_{\theta^*}(\mathbf{x}_{1:j(K} | y_{1:K})$ and the likelihood $p_{\theta^*}(y_{1:K})$.  \;
        \item Sample $X_{1:k}^* \sim \hat{p}_{\theta^*}(\cdot | y_{1:K})$ and the likelihood  $\hat{p}_{\theta^*}(y_{1:K})$ \;
        \item Accept proposed parameters and workload trajectories with probability
            $$
             \min\bigg(1, \frac{\hat{p}_{\theta^*}(y_{1:K})}{\hat{p}_{\theta(m-1)}(y_{1:K})} \cdot \frac{p(\theta^*)}{p(\theta(m-1))} \cdot \frac{q(\theta(m-1) | \theta^* )}{q(\theta^* | \theta(m-1))} \bigg).
            $$ 
            \item If accepted,  set $\theta(m)=\theta^*, X_{1:k}(m)= X_{1:k}^*$ and $ \hat{p}_{\theta(m)}(y_{1:K})=\hat{p}_{\theta^*}(y_{1:K})$. If not accepted, set $\theta(m)=\theta(m-1),  X_{1:k}(m)= X_{1:k}(m-1)$ and $ \hat{p}_{\theta(m)}(y_{1:K})=\hat{p}_{\theta(m-1)}(y_{1:K})$. \;
    \end{itemize}

}

} 

\KwOut{Posterior distribution over the parameters and workload trajectories $p( \mathbf{\theta},\mathbf{x}_{1:K} | y_{1:K})$ }

\caption{Particle Marginal Metropolis-Hastings (PMMH)}  \label{alg:pmmh}

\end{algorithm}

\begin{algorithm}[t]
\SetNlSty{textnormal}{}{}
\DontPrintSemicolon
    \caption{Particle Marginal Metropolis-Hastings}
    \label{alg:simple sampling}
    \KwIn{Initial $\theta_0$, a particle filter $\textsc{pf}$ (weight $g_{\epsilon}$), parameter sampler $q$, observations $y_{1:T}$, and a distribution $p$ on $\theta$.}
    Set $\textsc{pf}(y_{1:T} | \theta_0)$ to be a small positive number.\;
    \For{$i=0$ to $m-1$}{
        Generate $\vartheta \sim q(\theta | \theta_i)$\;
        Generate $U \sim \textnormal{Uniform}(0, 1)$\;
        
        $a = \displaystyle\frac{\textsc{pf}(y_{1:T} | \vartheta) p(\vartheta) q(\theta_i | \vartheta)}{\textsc{pf}(y_{1:T} | \theta_i) p(\theta) q(\vartheta | \theta_i)}$\;
        $\theta_{i+1} = 
            \begin{cases}
		        \vartheta  & \textnormal{if} \; U \leq a,  \\
		        \theta_i & \textnormal{otherwise}.
	        \end{cases}
	    $\;
	    $X_{1:T, i+1} \sim \textsc{pf}(\theta_{i+1} | y_{1:T})$\;
    }
    \Return $\{\theta_i, X_{1:T, i}\}_{i=1}^m$
\end{algorithm}

\section{Numerical Experiments} 

In this section we test the performance of the PMMH approach to jointly estimating parking occupancy and the model parameters. First, we validate the approach by estimating parking occupancy and the model parameters using simulated data from a $GI/GI/s$ queue assuming all customers pay for parking ($p=1$). Next, we jointly estimate the parking occupancy and the model parameters under partial payment compliance. Finally, we apply the method to parking meter payment data from SF\textit{park}, a large scale parking intervention policy in San Francisco.

\subsection{Simulated Data} \label{sec:results:simulation}

First, we assume that the drivers pay for parking 100\% of the time. In order to validate the estimation technique, we simulate 40 payments using the parking (Section \ref{sec:model:parking}) and payment (Section \ref{sec:model:parking}) model with parameters  $(\lambda, \mu, p)=(0.752, 5.0, 1.0)$ and with $7$ spaces. We perform the inference according to Algorithm \ref{alg:pmmh} by simulating a candidate set of state trajectories and parking model parameters $(\lambda, \mu)$. Next, we run a particle filter to generate $N=600,000$ samples of the state from the posterior distribution, and an estimate of the data likelihood. The proposed state is a sample from the $N$ particles and the model parameters. We accept or reject the proposal with a probability proportional to the estimated likelihood of the observed payments. We iterate this procedure until we accept a predefined number of proposals in excess of a burn-in period.  

\begin{figure}[htbp] 
\includegraphics[scale=0.5]{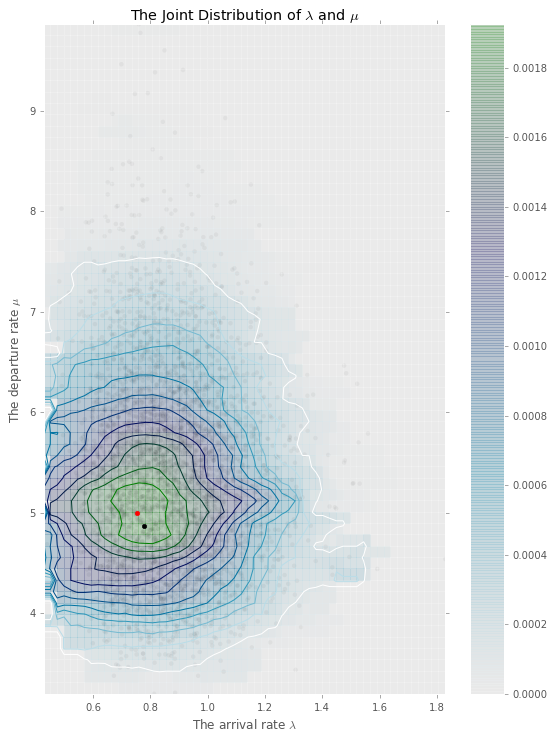}
\caption{Joint distribution of the model parameters accepted by the PMMH the under the model parameters  $(\lambda, \mu, p)=(0.752, 5.0, 1.0)$ and with $7$ spaces.  } \label{fixed_p_joint_lam_mu}
\end{figure}

\begin{figure}[th] 
\includegraphics[scale=.4]{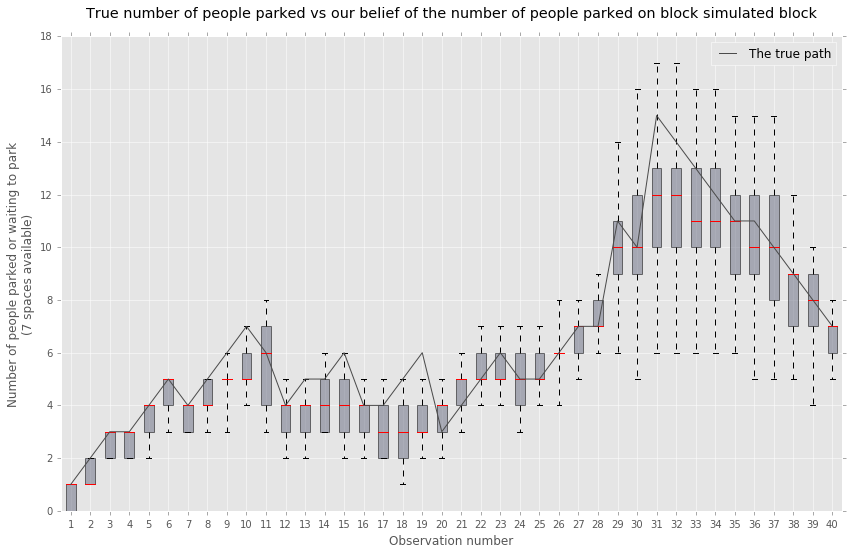}
\caption{The solid line is the parking occupancy trajectory at 40 simulated payments time under the model parameters  $(\lambda, \mu, p)=(0.752, 5.0, 1.0)$ and with $7$ spaces. The estimated parking occupancy trajectory using PMMH, with 600,000 particles, is shown as a box-stem plot at the payment times. The most likely value of the parameters from the estimation are $(\lambda, \mu, p)=(0.935, 5.195, 1.0)$ The root mean square difference of the median of our belief against the true path is 1.12 cars. 
} \label{fig:sim_fixed_p}
\end{figure}

The output of the PMMH is a joint posterior distribution of the model parameters and state trajectories. Figure \ref{fixed_p_joint_lam_mu} is the contour plot of the accepted model parameters $(\lambda,\mu$). From this joint distribution, we select the most likely parameter-trajectory pair, $(\bar{\lambda}, \bar{\mu}) = (0.935, 5.195 )$, with the largest value of the data likelihood.  Figure \ref{fig:sim_fixed_p} plots the mean and confidence intervals of accepted trajectories generated by the PMMH algorithm. The true path is computed from the parking queueing model corresponding to the $40$ observed payments. We find that the true trajectory falls well within the 95\% confidence interval of the estimated trajectory distribution. The corresponding root mean square error is 1.12 cars.

Next, we assume that 80\% of drivers pay for their parking while 20\% do not. We simulate a new sequence $40$ payments from a model with the same parameters $(\lambda, \mu, p) = ( 0.752,  5.0,     0.8  )$ and 7 parking spaces. From the generated contour plot of the accepted model parameters, we select the most probable  parameter-trajectory pair,  $(\bar{\lambda}, \bar{\mu}, p) = (0.891, 4.975, .808 )$, which has the largest value of the data likelihood.  Figure \ref{fig:sim_fixed_p} plots the mean and confidence intervals of accepted trajectories generated by the PMMH algorithm. The true path is computed from the parking queueing model corresponding with the $40$ observed payments. 

There is a noticed decrease in estimation accuracy caused by the increased number of model parameters. When $p <1$, the particle filter estimates the number of arriving drivers between consecutive payments. This trans-dimensional (see Green \cite{green1995reversible}) feature of the problem increases the number of variables that need to be inferred. This increases the complexity and decreases the accuracy of the estimator. The resulting root mean square error is 1.65 cars.

\begin{figure}[t] 
\includegraphics[scale=.45]{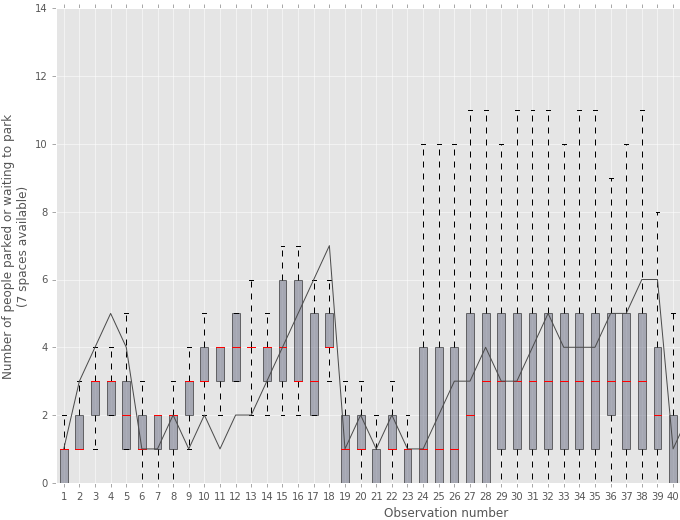}
\caption{ The solid line is the parking occupancy trajectory at 40 simulated payments time under the model parameters  $(\lambda, \mu, p)=(0.752, 5.0, 0.8)$ and with $7$ spaces. The estimated parking occupancy trajectory using PMMH, with 600,000 particles, is shown as a box-stem plot at the payment times. The most likely value of the parameters from the estimation are $(\lambda, \mu, p)=(0.891, 4.975, 0.808)$ The root mean square difference of the median of our belief against the true path is 1.65 cars.
} \label{floating_p}
\end{figure}

\subsection{Field Experiment: SF\textit{park}} \label{sec:results:sfpark}

We now apply the PMMH to SF\textit{park}, a large-scale smart parking initiative in the City of San Francisco. The goal is to reduce traffic by helping drivers find parking. SF\textit{park} uses real-time parking information and demand responsive pricing to achieve this goal. SF\textit{park}'s slogan is ``circle less and live more\footnote{http://SFpark.org/about-the-project/}." The program included more that on 7,000 on-street parking spaces.

We use two data sources in this experiment: occupancy snapshots and payment transactions. The payment transactions are inputs to the model. While the occupancy data set serves as ground truth to test the accuracy of the model predictions.  Many cities now have electronic parking payment meters/stations for on-street parking. These payment data are collected and stored in the memory of the parking meters. These parking data are commonly transmitted wirelessly from the parking meter to the vendor.

We use the same occupancy dataset as in Millard-Ball et al. \cite{millard2014curb}. We collected the occupancy data set by developing a web application that interacts with the SF\textit{park} API. These data provide snapshots of parking availability and capacity for each side of the street. The data contains occupancy snapshots at 5 minute intervals for 340 blocks. The observations are from January 1, 2012 to February 14, 2012. The API data set contains 1,730,770 observations during metered hours. 

The input dataset to the model consists of a payment data set that contains the date, time, block id, and payment amount for every transaction. One row of the payment has the form $\{'blockId':468022, 'Date':2012-01-03 T15:35:46.46, 'Amt': 1.25\}$. These payment data cover the same observation window as the occupancy data. There are over 3 millions payment transactions during the observation period January 1, 2012 to February 14, 2012.

\begin{figure}[t]
\includegraphics[scale=.42]{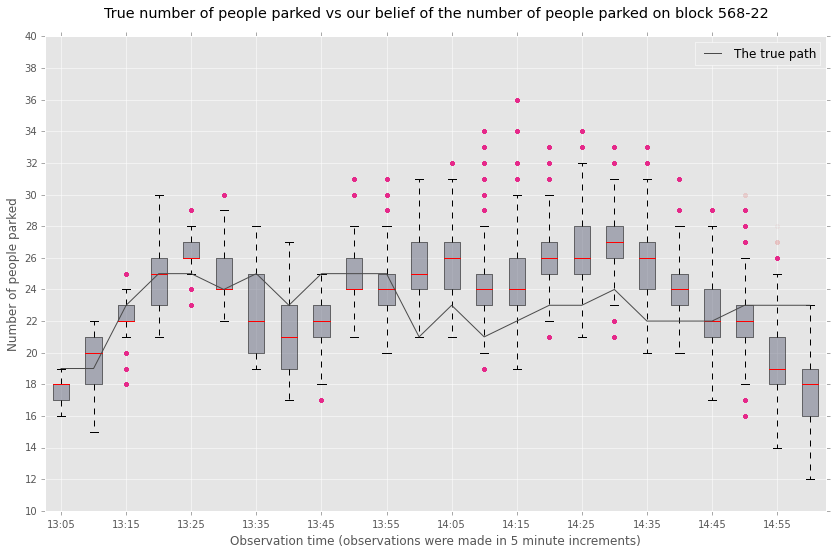}
\caption{The solid line is the parking occupancy trajectory for the 2200 block of Mission St. in San Francisco, CA on January 12, 2013 from 1pm-3pm. The estimated parking occupancy trajectory using PMMH, with 600,000 particles, is shown as a box-stem plot at the payment times.  The root mean square difference of the median of our belief against the true path is 2.54 cars. } \label{block_568-22}
\end{figure}

\begin{figure}[t] 
\includegraphics[scale=.42]{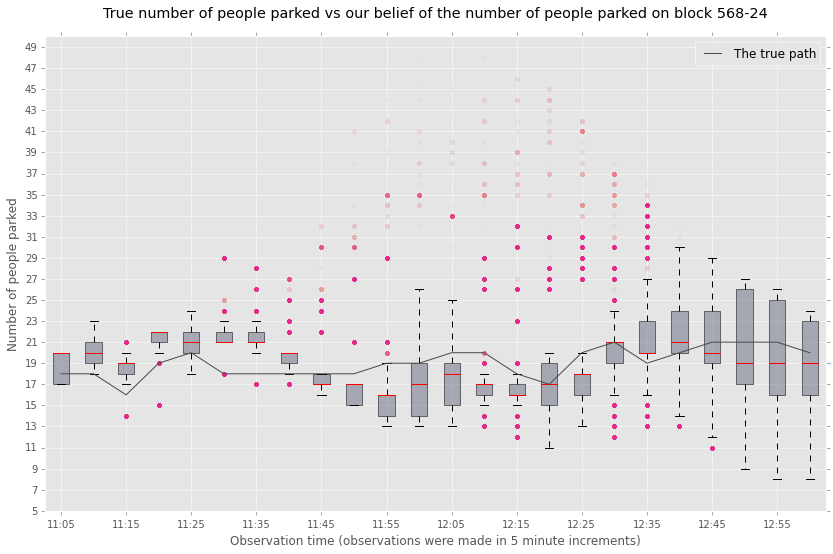}
\caption{ The solid line is the parking occupancy trajectory for the 2400 block of Mission St. in San Francisco, CA on January 18, 2013 from 11am-1pm. The estimated parking occupancy trajectory using PMMH, with 600,000 particles, is shown as a box-stem plot at the payment times.  The root mean square difference of the median of our belief against the true path is 2.01 cars.} \label{block_568-24}
\end{figure}

In this experiment, we use 160,000 particles for each run of the particle filter. In order to speed up the Metropolis-Hastings algorithm, we use an adaptive random walk mechanism to generate parameter proposals \cite{peters2010ecological}. The burn-in period of the adaptive Metropolis-Hastings step was 200 accepts, with 3,800 accepts after burn-in. We selected the 2200 and 2400 blocks of Mission Street in San Francisco for the inference. Figure \ref{block_568-22} shows a plot of the estimated and true occupancy trajectory over a 2 hours time window on the 2200 block of Mission Street in San Francisco on January 12, 2013. Figure \ref{block_568-24} shows a plot of the estimated and true occupancy trajectory over the 2 hours time window on the 2400 block of Mission Street in San Francisco on January 18, 2013. As expected, the mean squared error is the larger for the field test than the simulate data. However, the performance of our approach is superior to SF\textit{park}'s Sensor Independent Rate Adjustment (SIRA) regression approach.

\subsection{Sensor Independent Rate Adjustment Approach}

SF\textit{park} developed a method  to estimate the hourly parking occupancy rate from hourly payment rate data \cites{SIRA2014, demisch2016demand}. Occupancy rate is total occupied time divided by the sum of occupied and unoccupied time. The payment rate is the total paid time divided by the sum of the total paid time and unpaid time.  It is based on multiple linear regression, where dummy variables correspond to the various neighborhood in San Francisco. SF\textit{park} currently uses SIRA to generate monthly occupancy estimates. These occupancy estimates influence the monthly changes in parking prices.

Our method produces occupancy estimates at a finer time resolution than the SIRA method. In addition to parking occupancy, we estimate the arrival rates and parking times which SIRA does not do. We also ran test of our method against SF\textit{park} SIRA method. During the same time and location that we used above, the true hourly average occupancy was 84\%. Our method estimated the hourly average occupancy at 88\%, while SIRA estimated an hourly average occupancy of 105\%. 

\subsection{Implications for practice}
As cities and urban areas become more data rich, examples of data being re-purposed or given parallel lives will become more ubiquitous, we just demonstrated one. In addition, our numerical methods show that revisions to the estimates could be made in real time, thus allowing the dynamic pricing of parking spaces in a manner that reflect both the temporal and spatial dimension of the demand. We have demonstrated the effectiveness of our approach in achieving SF\textit{park}’s occupancy objective given the increased reliability in our estimates compared to SIRA’s. We have also shown how our approach could be replicated nationwide – a feature that delivers on a core objective of SF\textit{park} as a demonstration project. This could be achieved at minimal cost compared to the invasive and expensive price tags typically associated with in-ground sensors and the reduction in reliability when their batteries start to fail.

The ability to modify parking rates, particularly in real time, has implications for alleviating parking search related congestion. Although urban planners know of the areas where there is scarce parking, they do not have a good way of quantifying the magnitude of the problem. Once these neighborhoods are identified, we can look to a variety of remedies, such as: changing parking rates to incentivize drivers to choose off-street instead of searching for on-street parking; rezoning areas so that more off-street parking is available; increasing off-street parking with direct investments; or implementing congestion pricing in neighborhoods with high rates of cruising.

All four of the remedies mentioned above can be viewed as policies that either explicitly or implicitly change the price of parking. Changing the price of on-street parking is the most direct way to reduce drivers’ incentive to search for on-street parking and is the path that policy makers in SF\textit{park} took. Investing directly in off-street parking and rezoning for more off-street parking increases off-street parking supply and should reduce the price of off-street parking. These approaches are often difficult to implement in practice, and their effects are difficult to estimate beforehand. Using congestion pricing in high traffic areas functions as an implicit increase in the price of both on-street and off-street parking. Congestion pricing has other benefits since it reduces traffic overall, however, it is difficult to implement in a targeted manner.

\section{Conclusion}

This paper introduces an inference framework for estimating parking occupancy from parking meter payment data. Given a historical record of parking payment transactions, our method provides estimates of second-by-second parking occupancy. This framework allows cities to use existing infrastructure (parking meters) to estimate parking occupancy without installing specialized parking sensors. The inference methodology is an extension of a simulation based approach called Particle Markov Chain Monte Carlo method that includes an approximate Bayesian computation step to compute the likelihood weights. We show that this method yields unbiased asymptotic estimates of the posterior distribution.

The framework has the virtue of being unsupervised; that is,  no training set of ground truth parking occupancy data is required to perform the inference. The inference relies solely on the underlying model for the parking search process. Although we employed the $GI/GI/s$ queue with first-come first-serve, the PMMH framework supports any simulated parking search model. Beyond introducing the inference framework and establishing its basic theoretical properties, subsequent efforts may focus on developing the most appropriate parking search queuing model. This methodology can serve as the basis for a tool, that can potentially, enable a parking information system for citizens and avail policy makers and city planners the ability to evaluate the impact of parking policy interventions such as pricing modifications, information provision, and time limit changes.

\newpage

\end{document}